\documentclass[twocolumn]{aastex701}

\newcommand{\dmexc}{{\rm DM}_{\rm exc}}
\newcommand{\dmmw}{{\rm DM}_{\rm MW}}

\newcommand{\pccm}{{\rm pc\;cm}^{-3}}
\newcommand{\nside}{N_{\rm side}}
\newcommand{\nfrb}{N_{\rm FRB}}
\newcommand{\chime}{CHIME/FRB}

\shorttitle{Great Walls of Baryons}
\shortauthors{Ravi et al.}

\begin{document}

\title{Great Walls of Cosmic Baryons in the Northern Sky}

\author[orcid=0000-0002-7252-5485]{Vikram Ravi} 
\affiliation{Cahill Center for Astronomy and Astrophysics, MC 249-17, California Institute of Technology, Pasadena, CA 91125, USA}
\email{vikram@caltech.edu}
\author{Kritti Sharma}
\affiliation{Cahill Center for Astronomy and Astrophysics, MC 249-17, California Institute of Technology, Pasadena, CA 91125, USA}
\email{kritti@caltech.edu}
\author{Liam Connor}
\affiliation{Center for Astrophysics | Harvard \& Smithsonian, 60 Garden Street, Cambridge, MA 02138, USA}
\email{liam.connor@cfa.harvard.edu}


\begin{abstract}
The dispersion measures (DMs) of fast radio bursts (FRBs) encode the total ionized-gas column densities along their sightlines. Most observed FRBs originate at distances where the cosmological principle applies. Thus, variations in the DM distribution of FRBs observed in different regions on the sky trace local sources of anisotropy, such as the warm ionized medium and circum-galactic medium of the Milky Way, and local large-scale structure. We present a map of extragalactic DM variations across the Northern sky using a few thousand FRBs from the second \chime{} catalog. We detect a $\gtrsim 4\sigma$ excess of $\sim$150~$\pccm$ above the global mean, extended over $\sim$30$^\circ$ scales and centered near $\alpha \approx$~$12^{\rm h}$, $\delta \approx$~$55^\circ$. This excess, termed Wall~1, is robust to variations in sample selection and jackknife resampling, and cannot be explained by Galactic-disk DM-model uncertainties. The excess is likely too large to correspond to anisotropy in the Milky Way halo. The signal spatially coincides with the Ursa Major supercluster and associated large-scale structures. A secondary, more tentative Wall~2 near $\alpha \approx 2^{\rm h}$, $\delta \approx$~$45^\circ$ is spatially coincident with the Perseus-Pisces supercluster. Although the spatial coincidences suggest that the Walls may correspond to baryons in the local large-scale structure, the probability of chance coincidence is likely too high ($\sim10-20\%$) to claim confident associations. These results highlight the potential of using FRB DMs to detect baryon overdensities associated with local large-scale structure, and have important implications for near-field baryon mapping and FRB cosmology.
\end{abstract}


\section{Introduction} \label{sec:intro}

The large-scale structure of the local universe has been mapped in progressively greater detail since early galaxy redshift surveys revealed the filamentary cosmic web \citep{joeveer1978, gregory1978, delapparent1986, geller1989}. These pioneering observations established that galaxies are distributed in sheets and filaments surrounding large voids, with the densest concentrations forming galaxy clusters, which are themselves grouped into superclusters with total masses of $\sim10^{16}M_{\odot}$. Catalogs of rich galaxy clusters \citep{abell1958} enabled the identification of superclusters as coherent overdensities extending over tens to more than 100~Mpc \citep[e.g.,][]{bahcall1984, einasto2001}. Galaxy redshift surveys \citep[e.g., the 2MASS Redshift Survey][]{huchra2012} now provide a highly complete census of the galaxy distribution to a few hundred Mpc. X-ray selected supercluster samples from ROSAT \citep[e.g.,][ and references therein]{bohringer2021b} and from eROSITA \citep[e.g;,][]{liu2024} provide potentially more accurate mass estimates, and are now pushing the identification of superclusters to $z\sim0.8$.  

Although the galaxy distribution is widely used to trace the underlying matter-density field, locating the majority of cosmic baryons has proven more challenging. The total baryon density is precisely determined from Big Bang nucleosynthesis and the cosmic microwave background \citep{planck2020}, but only a small fraction of baryons are directly observed in stars, cold gas, and the intracluster medium \citep{fukugita1998, shull2012, connor2025}. X-ray observations probe hot ($T > 10^7$~K) gas in galaxy clusters but are insensitive to the more diffuse warm-hot intergalactic medium \citep{cen1999}. The thermal Sunyaev-Zel'dovich (SZ) effect has provided evidence for baryons in filamentary structures connecting clusters \citep[e.g.,][]{degraaff2019, tanimura2019, Li2026}. Absorption-line spectroscopy of background quasars probes the diffuse intergalactic medium but is limited to pencil-beam sightlines and requires large ionization corrections \citep{nicastro2018}. Inference on the baryon power spectrum \citep[e.g.,][]{siegel2025} based on X-ray and SZ tracers indicates a statistical suppression of structure on (sub-)megaparsec scales, providing evidence for the redistribution of matter from halos to the cosmic web caused by feedback. Major uncertainties remain in the spatial distribution of baryons, particularly at the interfaces between galaxy halos, the cosmic web, and the diffuse intergalactic medium (IGM).

Fast radio bursts \citep[FRBs;][]{lorimer2007, cordes2019, petroff2022} provide a powerful and complementary probe of cosmic baryons \citep[e.g.,][]{sharma2025,sharma2026}. The dispersion measure (DM) of each FRB encodes the total column density of free electrons along the line of sight, making FRBs sensitive to all ionized baryons regardless of temperature or density. The DM--redshift relation \citep[the ``Macquart relation'';][]{macquart2020} has been used to measure the total cosmic baryon density, consistent with the value from early-universe constraints. Recently, \citet{connor2025} used a large sample of FRBs localized by the 110-antenna Deep Synoptic Array (DSA-110) and other instruments to partition cosmic baryons between the IGM and galaxy halos, finding that $\sim$76\% of baryons reside in the diffuse IGM. Statistical analyses using large samples of unlocalized FRBs from \chime{} have yielded constraints on the circumgalactic medium of foreground galaxies \citep{connor2022, wu2023,kahinga2026} and evidence for correlations between FRB DMs and large-scale structure as traced by galaxies \citep{wang2025,shirasaki2026} and the SZ effect \citep{takahashi2025}. Similar results are found in detailed analyses of the sightlines of FRBs localized by the DSA-110 \citep{hussaini2025}. Direct, spatially resolved mapping of baryon overdensities using FRB DMs remains a key goal for upcoming large-sample surveys. 

In this paper, we use the second \chime{} catalog \citep{chimefrbcat2}, by far the largest FRB sample from a single survey to date, to construct a map of extragalactic DM variations across the northern sky. In Section~\ref{sec:mapping} we describe our data selection (Section~\ref{sec:data}), map-making procedure (Section~\ref{sec:mapmaking}), and robustness tests (Section~\ref{sec:robustness}). In Section~\ref{sec:associations} we cross-match the observed DM excess regions with known large-scale structures. We discuss our results and their implications in Section~\ref{sec:discussion}.

\section{A Sky Map of Excess DM with \chime{} Catalog 2} \label{sec:mapping}

\subsection{Data Selection} \label{sec:data}

The second \chime{} catalog \citep{chimefrbcat2} contains 4606 FRBs\footnote{This does not include sidelobe detections, and events that should be excluded from population analyses due to non-nominal telescope operation.} detected between July 2018 and September 2023 with the Canadian Hydrogen Intensity Mapping Experiment \citep[CHIME;][]{chimefrbcat1} at 400--800~MHz. These originate from 3524 unique sources, after removing repeating bursts and sub-bursts. For each of these events, we use the catalog to extract sky positions, DM measurements, and signal-to-noise ratios (S/Ns).

We apply the following quality cuts to construct our fiducial analysis sample. We later consider the impact of varying these quality cuts. First, we impose a minimum \texttt{bonsai} pipeline S/N threshold of ${\rm S/N} \geq 9$. This cut removes low-significance events whose astrophysical nature is uncertain; for example, the highest-DM event in the catalog, FRB\,20230501D (DM of 8878.4\,$\pccm$) falls below this threshold. Second, we exclude events within $10^\circ$ of the Galactic Plane ($|b| < 10^\circ$), where Galactic DM model uncertainties are largest. Third, we restrict our analysis to the Northern sky ($0^\circ \leq \delta \leq 90^\circ$), given the limited population and potential selection effects (see below) at lower declinations. Where the refined DM measurement ({\tt dm\_fitb}) is unavailable, we use the \texttt{bonsai} DM estimate ({\tt bonsai\_dm}); the two agree to within a median difference of $\sim0.5$~$\pccm$ where both are available. To derive the extragalactic DM ($\dmexc$), we subtract the Galactic DM contribution predicted by the recent NE2025 model \citep{ocker2026}. We compute the NE2025 Galactic DM at HEALPix \citep{gorski2005} pixel centers ($\nside = 16$) and interpolate to each FRB position. Events with $\dmexc \leq 0$ are removed. Our final sample comprises 2812 FRBs with a global mean $\langle \dmexc \rangle = 617$\,$\pccm$ and standard deviation $\sigma_{\rm DM} \approx$419\,$\pccm$.

\begin{figure*}[t!]
\centering
\includegraphics[width=\textwidth]{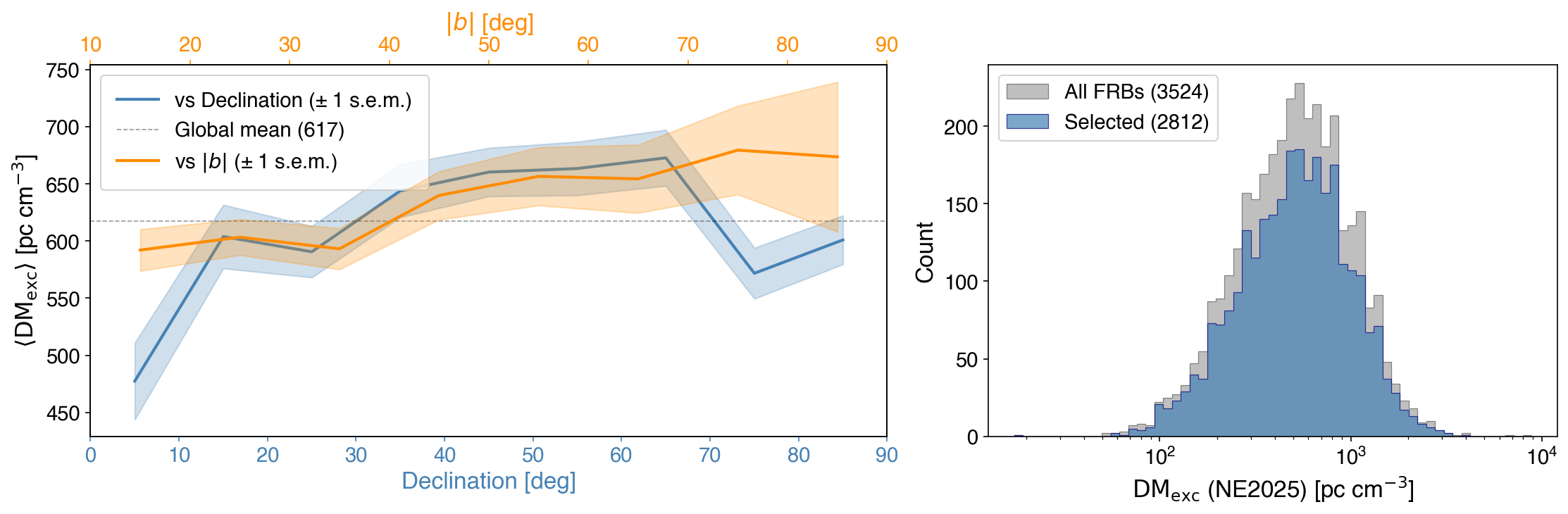}
\caption{{\it Left:} Mean extragalactic DM ($\langle \dmexc \rangle$, NE2025 model) as a function of declination (blue curve) and $|b|$ (orange curve), in $10^\circ$ bins. The shaded regions show the standard errors in the mean (s.e.m.) in each case. The horizontal dashed line marks the global mean (617~$\pccm$). Only marginal dependencies on declination and $|b|$ are observed, indicating that the sample is largely statistically homogeneous. {\it Right:} Distribution of $\dmexc$ (NE2025) for all FRBs after quality-flag and repeater cuts, and the exclusion of events with negative $\dmexc$ (gray; $N = 3492$) and the final selected sub-sample (blue; $N = 2812$).}
\label{fig:dmvsdec}
\end{figure*}

Figure~\ref{fig:dmvsdec} (left panel) shows  $\langle \dmexc \rangle$ as a function of declination (blue curve), and as a function of the absolute value of Galactic latitude ($|b|$, orange curve), both computed in $10^\circ$ bins. In both cases, given the errors in $\langle \dmexc \rangle$, only marginally significant trends are observed. The CHIME sensitivity is maximized at a declination around $\delta \sim+49^\circ$ \citep{chime2018}, which may be expected to result in a declination-dependent DM distribution \citep{shannon2018}. A drop in $\langle \dmexc \rangle$ at the lowest declinations ($\delta < 10^\circ$) is present but these bins are not central to our analysis, and our analysis below is robust to choosing a minimum declination of $+10^\circ$ rather than $0^\circ$. Similarly, given the sensitivity dependence of the \chime~pipeline on DM, a dependence of $\langle \dmexc \rangle$ on $|b|$ may be expected given the varying Galactic DM contribution. Only a tentative increasing trend in $\langle \dmexc \rangle$ with increasing $|b|$ is observed, and our analysis below is robust to variations in our minimum $|b|$ cutoff. The lack of a significant systematic dependence of $\langle \dmexc \rangle$ on declination or $|b|$ allows us to proceed with our map-making below without correcting for associated selection functions. The right panel of Figure~\ref{fig:dmvsdec} shows the DM distribution of the full sample after quality-flag and repeater cuts and of the final selected sample.

\subsection{Map Making} \label{sec:mapmaking}

Our goal is to identify statistically significant spatial variations in $\langle \dmexc \rangle$ across the sky. We approach this in two stages: an unsmoothed per-pixel map to identify spatially correlated structure, followed by a Gaussian-smoothed map to optimize signal detection.

\begin{figure*}[t!]
\centering
\includegraphics[width=\textwidth]{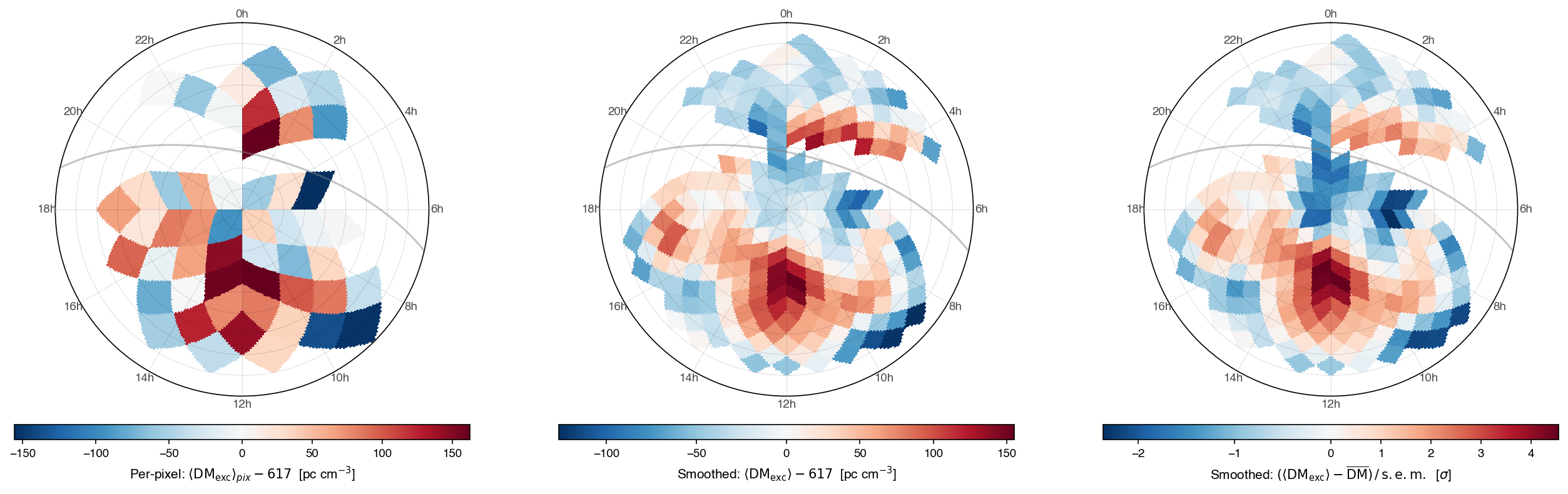}
\caption{North Celestial Pole--centered polar projections of extragalactic DM excess across the northern sky. RA increases clockwise; the gray curve marks the Galactic plane ($b = 0^\circ$). {\it Left:} Per-pixel mean DM excess ($\nside = 4$, $\nfrb \geq 20$ per pixel, 75 pixels). {\it Middle:} Gaussian-smoothed DM excess ($\nside = 8$, $\sigma = 7.3^\circ$). A prominent excess, which we term Wall~1, of $\sim$150~$\pccm$ is visible near $\alpha \approx$~$12^{\rm h}$, $\delta \approx$~$55^\circ$. A secondary excess, termed Wall~2, is visible near $\alpha \approx 2^{\rm h}$, $\delta \approx 45^\circ$. {\it Right:} Significance of the smoothed DM excess in units of the standard error of the mean. The peak exceeds $4\sigma$ for Wall~1. In all panels, the colorbars ranges are set by the minima and maxima of the data.}
\label{fig:dmmaps}
\end{figure*}

For the per-pixel map, we assign each FRB to a HEALPix pixel at $\nside = 4$ (pixel scale $\approx 14.7^\circ$), compute $\langle \dmexc \rangle$ in each pixel, and mask pixels with fewer than 20 FRBs. This yields 75 pixels covering the northern sky above $|b| > 10^\circ$. The left panel of Figure~\ref{fig:dmmaps} displays the per-pixel DM excess, defined as $\langle \dmexc \rangle_{\rm pix} - 617$~$\pccm$. Even at this coarse resolution, clear spatial correlations are visible: a contiguous region of excess DM near $\alpha \approx 11^{\rm h}$--$14^{\rm h}$, $\delta \approx 30^\circ$--$60^\circ$.

To better characterize the signal, we construct a Gaussian-smoothed map at $\nside = 8$ (pixel scale $\approx 7.3^\circ$). For each pixel center, all FRBs within $3\sigma$ angular distance ($\sigma = 7.3\circ$) are included with Gaussian weights $w = \exp[-\theta^2 / (2\sigma^2)]$, where $\theta$ is the angular separation. The weighted mean DM is computed, and an effective number of independent samples is estimated as $n_{\rm eff} = (\sum w)^2 / \sum w^2$. Pixels are masked where the number of FRBs within $2\sigma$ is fewer than 50. The middle panel of Figure~\ref{fig:dmmaps} shows the smoothed DM excess, revealing a prominent positive feature centered near $\alpha \approx 12^{\rm h}$, $\delta \approx 55^\circ$ with an amplitude of $\sim$150~$\pccm$. The very approximate extent of this excess, from the per-pixel map, is over $\sim$30$^\circ$. The right panel shows the corresponding significance map, $(\langle \dmexc \rangle_{\rm smooth} - 617) / {\rm s.e.m.}$, where s.e.m. is the standard error in the mean, with the peak at $4.3\sigma$. It is possible that this feature is linked to other features at a similar declination, although these extensions are at lower significance. A second positive feature, reaching $\sim2.3\sigma$ significance, is present near $\alpha \approx 2^{\rm h}$, $\delta \approx 45^\circ$. We term these features Wall~1 and Wall~2 respectively (Table~\ref{tab:dm_excess}).

\begin{deluxetable*}{lccccc}
\tablecaption{Dispersion Measure Excesses in the Northern Sky\label{tab:dm_excess}}
\tablewidth{0pt}
\tablehead{
\colhead{Name} & \colhead{R.A.} & \colhead{Decl.} & \colhead{Extent} & \colhead{DM Excess} & \colhead{Peak significance} \\
\colhead{} & \colhead{(J2000)} & \colhead{(J2000)} & \colhead{(deg.)} & \colhead{(pc\,cm$^{-3}$)} & 
}
\startdata
Wall 1 & 12$^{\rm h}$ & $+$55\arcdeg & $\sim30$ & $\sim150$ & $4.3\sigma$ \\
Wall 2 & 02$^{\rm h}$ & $+$45\arcdeg & $\sim20$ & $\sim150$ & $2.3\sigma$ \\
\enddata
\end{deluxetable*}

The effective angular resolution of this analysis is fundamentally set not by the smoothing kernel alone but by the angular scale over which sufficient FRB counts accumulate to yield a statistically meaningful measurement of $\langle \dmexc \rangle$. Given the sample density, the effective resolution is $\sim$20$^\circ$. Structure on larger angular scales is revealed by the smoothing.

\subsection{Robustness Tests} \label{sec:robustness}

We subject the DM maps to extensive robustness tests. First, we examined sensitivity to data selection cuts. Increasing the minimum S/N threshold and using \texttt{fitburst} S/N thresholds, reducing the maximum allowed DM (to catch more potential outliers like FRB\,20230501D), varying the Galactic latitude cut, restricting to different date ranges, and including or excluding the catalog's ``excluded'' events all yield maps with consistent spatial structure. The primary effect of more restrictive cuts is a reduction in statistical significance, as expected from the smaller sample sizes, rather than a qualitative change in the morphology.

\begin{figure*}[t!]
\centering
\includegraphics[width=\textwidth]{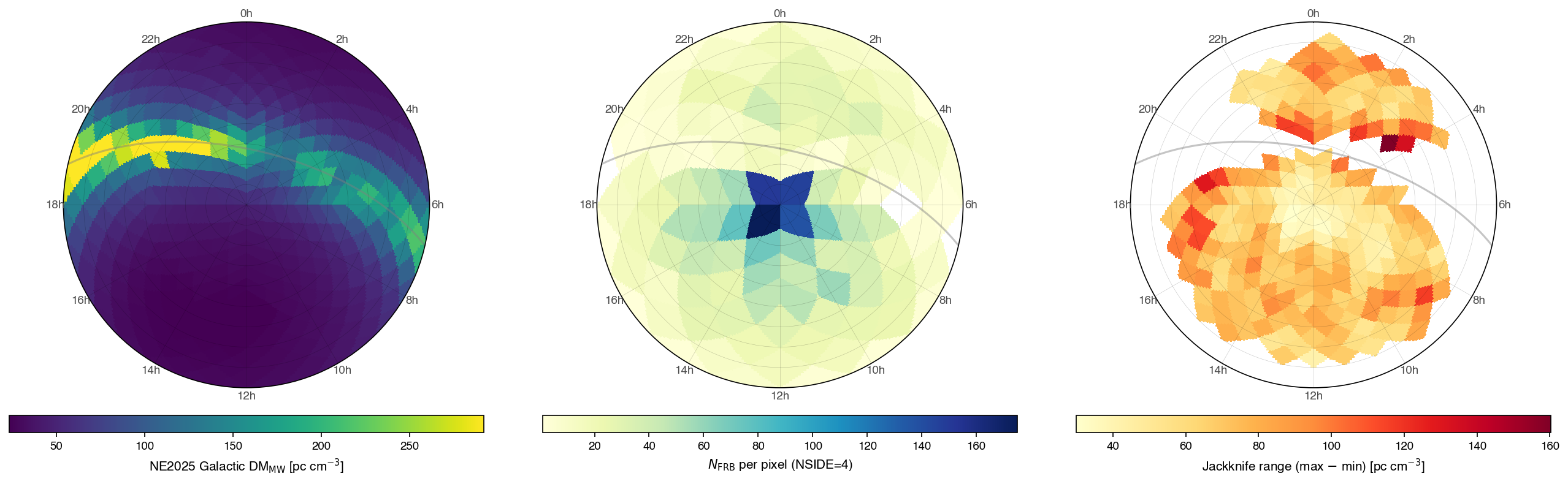}
\caption{Diagnostic maps. {\it Left:} NE2025 Galactic DM ($\dmmw$) on the northern sky ($\nside = 16$), showing values of 50--250~$\pccm$ with the highest contributions near the Galactic plane. {\it Middle:} Counts of all 2812 selected FRBs in $\nside = 4$ HEALPix pixels. The distribution is roughly uniform in RA, with a concentration toward the NCP reflecting CHIME's higher exposure at more northern declinations as a transit telescope. {\it Right:} Jackknife range (max $-$ min of DM contrast over 100 realizations each dropping 10\% of FRBs). Typical ranges of $\lesssim80~\pccm$ around Wall~1 confirms that this excess is robust to the removal of individual events, although a somewhat increased range in the region of Wall~2 points to a lower significance of this feature.}
\label{fig:diagnostics}
\end{figure*}

To formalize this, we performed jackknife resampling with 100 realizations, each created by randomly dropping 10\% of the FRBs. The right panel of Figure~\ref{fig:diagnostics} shows the jackknife range (maximum minus minimum of the DM contrast across all realizations) at each pixel. The range is typically $\lesssim70~\pccm$, confirming that the Wall~1 excess is robust and not driven by individual outlier events. A modestly elevated jackknife range is visible around the Wall~2 region, suggesting that this secondary feature there may be more sensitive to individual FRBs, consistent with its lower overall significance.

We also performed significance tests where we randomized critical FRB properties, and saw no reduction in the significance of the Wall features. We created 1000 realizations of the dataset with randomly drawn RAs \citep[e.g., following][]{rafiei2021,wang2025}, and 1000 realizations of the dataset with randomly drawn Galactic longitudes, and in both cases computed noise estimates in the Gaussian-smoothed maps for each pixel. The former exercise was intended to test the effect of a declination dependence in $\langle \dmexc \rangle$, and the latter was intended to test the effect of a Galactic-latitude dependence in $\langle \dmexc \rangle$ (see Figure~\ref{fig:dmvsdec} and related discussion). The significances of both Wall features were retained. Similarly, we also tested the robustness of the Wall features to scrambled values of the extragalactic DMs, and found consistent results. 

The spatial coherence of the detected signals---extended, contiguous regions of excess DM spanning multiple independent HEALPix pixels--- argues against noise fluctuations or isolated systematic artifacts. Purely random variations would not produce the degree of spatial correlation visible in the maps. Indeed, comparing the Moran's $I$ spatial coherence statistic\footnote{The $I$ statistic is computed as the normalized ratio of the spatial covariance of adjacent pixels to the global variance.} of the real per-pixel map with 1000 realizations of maps made from position shuffles of the FRBs indicates that the real data have a degree of coherence that has a $<1.6\%$ chance of being due to chance. A similar test where the RAs of the events are randomized yields a $<3.0\%$ chance probability of obtaining the observed spatial coherence. The middle panel of Figure~\ref{fig:diagnostics} shows the sky distribution of the counts of selected FRBs, confirming that the Walls are not consistent with over- or under-densities in the FRB detections.

Finally, we tested robustness of the Walls to the Galactic DM model. Replacing NE2025 with NE2001 \citep{ne2001} or YMW16 \citep{ymw16} produces no significant change in the DM excess morphology. This is expected: the left panel of Figure~\ref{fig:diagnostics} shows that the NE2025 total Galactic DM around Wall~1 is typically $\lesssim50~\pccm$, far smaller than the $\sim$150~$\pccm$ excess DM. Variations between models are smaller still. The NE2025 Galactic DM around Wall~2 is larger ($\sim100~\pccm$), but the error in the model would have to more than double this to explain the observed excess DM. The Milky Way circumgalactic medium \citep[CGM;][]{cook2023, ravi2025} likely contributes $\sim30~\pccm$ in total, with few tens of percent variations from sightline to sightline \citep[e.g.,][]{foh2023}. Thus, we exclude a Galactic origin for the observed  Wall~1 and Wall~2 features.

\section{Associations with Known Superclusters} \label{sec:associations}

The extragalactic DM along a given sightline receives contributions from baryons at all redshifts between the source and the observer. However, the cosmological principle implies that the DM excesses we observe in Wall~1 and Wall~2 originate from relatively nearby, extended baryon overdensities. We now examine whether the detected features correspond to known structures in the local universe. 

Although predicted DM contributions from specific extragalactic structures are uncertain, observations are beginning to yield positive detections. \citet{prochaska2019} and \citet{ocker2022} modeled the DM distribution in the local universe using galaxy density reconstructions (see also \citet{konietzka2025}), and predict anisotropies at the level of tens of $\pccm$ due to local halos and filaments. In a comparable work to ours, \citet{liu2026} found tentative evidence for a dipole anisotropy in the Galactic halo DM, with a peak excess (in their refined sample) that coincides within the large respective error ranges with our Wall~1. \citet{connor2022} and \citet{wu2023} estimated DM contributions from foreground galaxy halos using \chime~Catalog 1 data, and found associated excesses on the order of $100~\pccm$ within the virial radii. It has proven difficult to associate DM excesses with specific structures smaller than massive galaxy groups and clusters \citep[e.g.,][]{prochaskaFRB2019}, likely because smaller structures are baryon deficient \citep{ravi2025,connor2025}, although see \citet{kahinga2026} for a potential excess DM associated with the M31 halo in \chime~Catalog 2. FRBs originating within galaxy clusters clearly exhibit measurable DM excesses \citep{connor2023,lanman2025}.

\begin{figure*}[t!]
\centering
\includegraphics[width=0.75\textwidth]{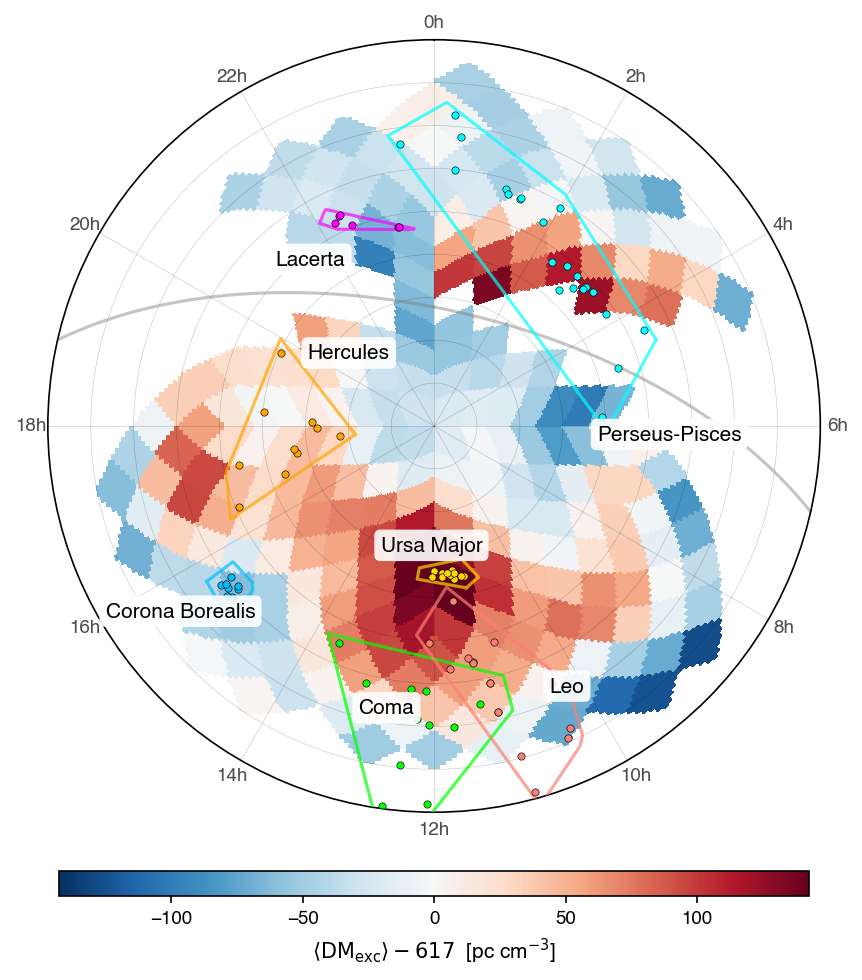}
\caption{Gaussian-smoothed DM excess map (as in the middle panel of Figure~\ref{fig:dmmaps}) with nearby supercluster positions overlaid. Colored circles and convex-hull outlines show member clusters from four catalogs: CLASSIX superclusters \citep{bohringer2021c} — Perseus-Pisces (cyan), Coma (lime), Hercules (orange), and Lacerta (magenta); the Ursa Major supercluster from \citet{kopylova2009} (gold); the Leo supercluster from \citet{kopylova2011} (salmon); and the Corona Borealis supercluster from \citet{pearson2014} (sky blue). All labels are placed adjacent to their respective convex-hull outlines. The gray curve marks the Galactic plane. The Wall~1 feature coincides with the Ursa Major supercluster, and the more tentative Wall~2 feature coincides with the Perseus-Pisces supercluster.}
\label{fig:overlay}
\end{figure*}

The angular scales, and DM-excess amplitudes of the Wall features severely restrict the structures under consideration. Neither M31 nor the M81 group are spatially coincident with the Walls. The very nearest galaxy cluster in the north (Virgo) is also not coincident, nor are other very nearby clusters. We therefore turn to the nearest superclusters: groups of two or more galaxy clusters representing an overdensity, but not necessarily gravitationally bound \citep{bahcall1988}. In Figure~\ref{fig:overlay}, we show the positions of the seven nearest northern superclusters; in approximate order of distance, these are: Perseus-Pisces \citep{bohringer2021b}, Lacerta, Coma, Hercules \citep{bohringer2021c}, Leo \citep{kopylova2011}, Ursa Major \citep{kopylova2009}, Corona Borealis \citep{pearson2014}.  Figure~\ref{fig:overlay} overlays the positions of Abell galaxy clusters that are members of these superclusters, along with convex-hull outlines around these positions, on the smoothed DM excess map. We highlight two notable associations.

\begin{enumerate}

\item {\it Ursa Major.} The Ursa Major supercluster appears spatially associated with the Wall~1 feature. Ursa Major is a moderately rich, filament-type supercluster at $z\sim0.06$, comprising $\sim10$ Abell clusters arranged along three filamentary substructures with a total dynamical mass of $\sim 2.3 \times 10^{15}M_{\odot}$ \citep{kopylova2009}. The system is largely at rest in the Hubble flow \citep{kopylova2007}, and overdensity estimates suggest that it is gravitationally bound. Its composite luminosity function resembles that of the field rather than dynamically evolved superclusters like Corona Borealis, and its member clusters show unusually tight scaling relations suggestive of a common formation history \citep{kopylova2009}. The Ursa Major supercluster is relatively isolated, with no rich X-ray clusters or neighboring superclusters in its vicinity \citep{kopylova2009, krause2013}.

\item {\it Perseus-Pisces.} The Perseus-Pisces supercluster overlaps with the more tentative Wall~2 feature. The Perseus-Pisces supercluster is one of the most prominent structures in the nearby universe, forming a dense, elongated chain of galaxies and clusters spanning $\sim100$\,Mpc in length and $\sim12$\,Mpc in width at $z \approx 0.016$, anchored by the massive Perseus cluster (A426) along with A262 and A347 \citep{joeveer1978}. Its remarkably coherent filamentary morphology, lying nearly in the plane of the sky, made it one of the first structures to demonstrate the topology of the cosmic web \citep{Giovanelli1986}. Using X-ray-selected cluster data, \citet{bohringer2021b} established Perseus-Pisces as the largest superstructure at redshifts $z\lesssim0.03$, with a total estimated mass of $\sim1.3 \times 10^{16}M_\odot$. Unlike more isolated systems such as the Ursa Major supercluster, Perseus-Pisces is embedded in a rich large-scale environment, with member clusters showing X-ray luminosities systematically higher than field clusters \citep{bohringer2021b}.

\end{enumerate}

A well-motivated physical interpretation of these associations is not immediately evident. The Wall~1 feature is likely spatially extended relative to the Ursa Major supercluster, suggesting that if the association is true, baryons associated with the supercluster are distributed throughout the surrounding large-scale structure. It is not clear why baryon excesses appear to be more likely associated with some superclusters, and not with others, although we note that Wall~1 appears tentatively extended towards Hercules, and may overlap with Coma and Leo. The diffuse baryon structures (in terms of content,  patchiness, and geometry) associated with specific superclusters are not possible to predict with high accuracy, and this uncertainty must be convolved with the uncertainty in the FRB sampling of the sky. Given an angular diameter $\theta$, a peak associated DM excess of ${\rm DM}_{\rm Wall}$, and a distance $D_{\rm Wall,\,Mpc}$, a spherical-cow estimate of the baryonic mass of a Wall is 
\begin{equation}
    M_{\rm Wall} \approx 1.3\times10^{10} M_{\odot} \times {\rm DM}_{\rm Wall} \theta^{2}  D_{\rm Wall,\,Mpc}^2,
\end{equation}
assuming a hydrogen plasma. Based on the data in Table~\ref{tab:dm_excess}, and assuming distances to the Walls corresponding to the potentially associated supercluster distances, we find baryon masses for Walls 1 and 2 of $\sim2\times10^{16}M_{\odot}$ and $\sim1\times10^{15}M_{\odot}$. Although the mass implied for Wall~2 is plausible, the mass implied for Wall~1 is implausibly high relative to the estimated dynamical mass of the Ursa Major supercluster. This may indicate the superposition or combination of multiple cosmic-web structures, and/or a significantly different geometry and physical extent. Detailed follow-up analyses of the range of possible DM excesses associated with local large-scale structures are required to aid in the interpretation of our results.  

We emphasize that these associations are only tentative. The angular resolution of the present analysis is comparable to the angular extents of the superclusters themselves, limiting our ability to isolate contributions from individual structures. The covering fraction of the superclusters on the sky is very large, extending over $\sim13\%$ of the northern sky; this is a rough estimate of the chance alignment probability between the Walls and any supercluster. Clearly, far larger FRB samples are required to further extend the near-field mapping of cosmic baryons, primarily by improving the angular resolution of the maps. 

\section{Discussion and Conclusions} \label{sec:discussion}

We have presented a map of extragalactic DM variations across the northern sky, using 2812 FRBs from the second \chime~catalog \citep{chimefrbcat2}. Our main findings are as follows:
\begin{enumerate}

\item By comparing the mean DM of FRBs in different regions in the sky with the global mean, we identify spatially extended regions where FRBs have excess mean DMs. A prominent region, which we term Wall~1, is centered on $\alpha \approx 12^{\rm h}$, $\delta \approx 55^\circ$ and has an amplitude of $\sim$150~$\pccm$. A secondary, lower-significance region with similar amplitude, termed Wall~2, is located near $\alpha \approx 2^{\rm h}$, $\delta \approx 45^\circ$. The structures likely extend over a few tens of degrees on the sky. Wall~1 appears to have somewhat larger extensions, although these are of low significance.  

\item We perform a series of tests to assess the robustness of these detections. The Wall~1 feature in particular has a detectable spatial extension, revealed by a significant degree of spatial correlation between adjacent pixels in the mapping. The Wall features are robust to how the FRB data are selected, such as in declination and Galactic latitude, in date range, in minimum signal to noise ratio, and indeed are robust to jackknife resampling. We note however that jackknife resampling suggests that the Wall~2 feature is potentially dominated by a few events, making its veracity more tentative. The features are robust to the choice of model for DM contributed by the Galactic disk, and are too strong to be explained by uncertainties either in the disk or halo DM contributions. We also show that selection effects in declination and Galactic latitude in the \chime{} sample do not require correction. 

\item Given their spatial extents, and the limited spatial resolution of our analysis, we consider the possibility of associations with nearby superclusters of galaxies. We identify a spatial coincidence between Wall~1 and the Ursa Major supercluster, and a more tentative coincidence between Wall~2 and the Perseus-Pisces supercluster. We caution that these associations are tentative ($\sim10-20\%$ false-association probability). 

\end{enumerate}

Several limitations of the present analysis should be noted. First, our reliance on mean DM as a statistical estimator means that the significance of the detection is fundamentally limited by sample size. Many more FRBs are needed to improve the angular resolution and statistical power of such maps. Second, we are limited by our knowledge of asymmetries in the Milky Way's CGM, and indeed in the intra-group medium of the Local Group. Third, the associations with specific superclusters remain tentative. A more quantitative analysis would benefit from comparison with detailed models for the local cosmic web \citep[e.g.,][]{courtois2023}, or from constrained analyses that use the known galaxy distribution as a prior on the DM field.

Looking forward, FRBs offer unique promise for near-field baryon mapping with much larger samples. Future catalogs from \chime{}, the Canadian Hydrogen Observatory and Radio-transient Detector \citep[CHORD;][]{chord2019}, and the full Deep Synoptic Array \citep{hallinan2019} are expected to increase sample sizes by orders of magnitude, dramatically improving angular resolution and enabling tomographic reconstruction of the local baryon distribution. Synergies with multiple probes---including X-ray observations of the intracluster medium, SZ measurements of the warm-hot intergalactic medium, and quasar absorption-line studies of the diffuse IGM---will be essential for constructing a complete picture of the local baryon distribution.

This work has direct implications for cosmological inference with FRB samples. The initial motivation for this study was a search for the analog of the CMB peculiar-velocity dipole in FRB DMs. The detected DM variations of $\gtrsim100~\pccm$ are over an order of magnitude larger than the $\sim{\rm few}~\pccm$ expected from a cosmological dipole, which is what prompted the baryon-mapping exercise presented here. Any measurement of large-scale DM anisotropy---whether for constraining the cosmic dipole, measuring $H_0$, or probing the the baryon power spectrum via variance in the Macquart relation---must ultimately account for the local, spatially varying DM contribution from nearby large-scale structure. This systematic is effectively a varying and uncertain ``Milky Way--associated'' DM that may exceed the Galactic disk contribution in many directions, representing a significant contamination for combining FRB samples across the sky for precision cosmology.

\begin{acknowledgments}
We thank Kaitlyn Shin for extensive and very useful discussions on this work. This work was supported in part by an Alfred P. Sloan Research Fellowship. This material is based upon work supported in part by the National Science Foundation under CAREER Grant Number 2240032. 
\end{acknowledgments}

\facilities{CHIME}

\software{
{\tt numpy},
{\tt matplotlib},
{\tt healpy} \citep{gorski2005},
{\tt astropy} \citep{astropy2022},
{\tt mwprop} \citep{ocker2026},
{\tt Claude}
}

\bibliography{references}{}
\bibliographystyle{aasjournal}

\end{document}